\documentclass[aps,prl,twocolumn,superscriptaddress]{revtex4-1}

\usepackage{graphicx} 
\usepackage{amssymb} 
\usepackage{amsmath}
\usepackage{array}
\usepackage{tabularx}

\begin{document}

\title{Unconventional field-induced spin gap in an \textit{S} = 1/2 chiral staggered chain} 

\author{J. Liu}
\affiliation{Department of Physics, Clarendon Laboratory, University of Oxford, Parks Road, Oxford OX1~3PU, UK}

\author{S. Kittaka}
\affiliation{Institute for Solid State Physics, University of Tokyo, Kashiwa, Chiba 277-8581, Japan}

\author{R. D. Johnson}
\affiliation{Department of Physics, Clarendon Laboratory, University of Oxford, Parks Road, Oxford OX1~3PU, UK}

\author{T. Lancaster}
\affiliation{Centre for Materials Physics, Durham University, South Road, Durham DH1~3LE, UK}

\author{J. Singleton}
\affiliation{National High Magnetic Field Laboratory, Los Alamos National Laboratory, MS-E536, Los Alamos, NM 87545, USA}

\author{T. Sakakibara}
\affiliation{Institute for Solid State Physics, University of Tokyo, Kashiwa, Chiba 277-8581, Japan}

\author{Y. Kohama}
\affiliation{Institute for Solid State Physics, University of Tokyo, Kashiwa, Chiba 277-8581, Japan}

\author{J. van Tol}
\affiliation{National High Magnetic Field Laboratory, Florida State University, Tallahassee, Florida 32310, USA}

\author{A. Ardavan}
\affiliation{Department of Physics, Clarendon Laboratory, University of Oxford, Parks Road, Oxford OX1~3PU, UK}

\author{B. H. Williams}
\affiliation{Department of Physics, Clarendon Laboratory, University of Oxford, Parks Road, Oxford OX1~3PU, UK}

\author{S. J. Blundell}
\affiliation{Department of Physics, Clarendon Laboratory, University of Oxford, Parks Road, Oxford OX1~3PU, UK}

\author{Z. E. Manson}
\affiliation{Department of Chemistry and Biochemistry, Eastern Washington University, Cheney, WA 99004, USA}

\author{J. L. Manson}
\email{jmanson@ewu.edu}
\affiliation{Department of Chemistry and Biochemistry, Eastern Washington University, Cheney, WA 99004, USA}

\author{P. A. Goddard}
\email{p.goddard@warwick.ac.uk}
\affiliation{Department of Physics, University of Warwick, Gibbet Hill Road, Coventry, CV4~7AL, UK}

\begin{abstract}
We investigate the low-temperature magnetic properties of the molecule-based chiral spin chain [Cu(pym)(H$_2$O)$_4$]SiF$_6\cdot$H$_2$O (pym = pyrimidine). Electron-spin resonance, magnetometry and heat capacity measurements reveal the presence of staggered $g$ tensors, a rich low-temperature excitation spectrum, a staggered susceptibility and a spin gap that opens on the application of a magnetic field. These phenomena are reminiscent of those previously observed in non-chiral staggered chains, which are explicable within the sine-Gordon quantum-field theory. In the present case, however, although the sine-Gordon model accounts well for the form of the temperature-dependence of the heat capacity, the size of the gap and its measured linear field dependence do not fit with the sine-Gordon theory as it stands. We propose that the differences arise due to additional terms in the Hamiltonian resulting from the chiral structure of [Cu(pym)(H$_2$O)$_4$]SiF$_6\cdot$H$_2$O, particularly a uniform Dzyaloshinskii-Moriya coupling and a four-fold periodic staggered field. 

\end{abstract}
\maketitle

%%%%%%%%%%%%%%%%%%%%%%%%%%%%

Quantum phase transitions can be driven between gapped and gapless phases by applied magnetic field. This was demonstrated in $S = 1$ quasi-one dimensional (Q1D) chains~\cite{Hagiwara2006} and $S = 1/2$ two-leg ladders~\cite{RueggPRL2008,KlanjsekPRL2008,JeongPRL2013}. These systems possess a spin gap that is closed by an external field, leading to a transition to a gapless phase which can be described by the Tomonaga-Luttinger liquid (TLL) theory~\cite{Giamarchi2003}. In contrast, the excitation spectrum of $S = 1/2$ antiferromagnetic (AF) Heisenberg chains with uniform nearest-neighbor interactions is gapless up to the saturation field. However, this ground state is highly sensitive to small modifications. The dramatic effect of alternating local spin environments was first discovered through high-field neutron scattering and heat capacity experiments on the $S = 1/2$ AF staggered chain, Cu-benzoate [Cu(C$_6$D$_5$COO)$_2\cdot$3D$_2$O]~\cite{DenderPRL1997}. Here, the presence of alternating $g$ tensors as well as Dzyaloshinskii-Moriya (DM) interactions produce an internal staggered field perpendicular to the external field. The staggered field breaks rotational symmetry around the applied field, leading to a field-induced spin gap and anisotropic staggered susceptibility~\cite{OshikawaPRL1997a,AffleckPRB1999}. This gapped phase can be described using the sine-Gordon (SG) quantum-field theory~\cite{AffleckPRB1999} and a complex excitation spectrum is predicted including solitons, antisolitons and soliton-antisoliton bound states called breathers~\cite{EsslerPRB1998,AffleckPRB1999,EsslerPRB2003}. Such excitations were experimentally confirmed by electron spin resonance (ESR) in the SG spin chain [pym-Cu(NO$_3$)$_2$(H$_2$O)$_2$] (pym = pyrimidine)~\cite{FeyerhermJPhysCondMatt2000,ZvyaginPRL2004}. So far these studies have been limited to $g$ tensors that alternate with a two-fold periodicity along the chain. 

In this letter, we report a detailed investigation of the chiral chain [Cu(pym)(H$_2$O)$_4$]SiF$_6\cdot$H$_2$O, in which adjacent Cu(II) environments are related by $4_1$ screw symmetry~\cite{CordesCrysGrowDesign2007}. At zero field, the magnetism of [Cu(pym)(H$_2$O)$_4$]SiF$_6\cdot$H$_2$O is well described by an $S = 1/2$ chain model and no long-range order is detected above 0.02~K, as evidenced by muon-spin relaxation measurements~\cite{SI}. Single-crystal ESR shows the presence of staggered $g$-tensors, anisotropic staggered susceptibility and a rich low-temperature excitation spectrum. However, the low-temperature field-induced excitation gap observed in heat capacity is significantly suppressed compared to that of a traditional non-chiral staggered chain, and has an unexpected linear field dependence. We propose this is due to additional interactions arising from the four-fold periodic chiral structure. This study showcases how the ground state of a Q1D AF chain is modified when competing interactions are introduced and poses a challenge to theorists to develop new models to fully account for the data. 

\begin{figure}[t]
\centering
\includegraphics[width=\columnwidth]{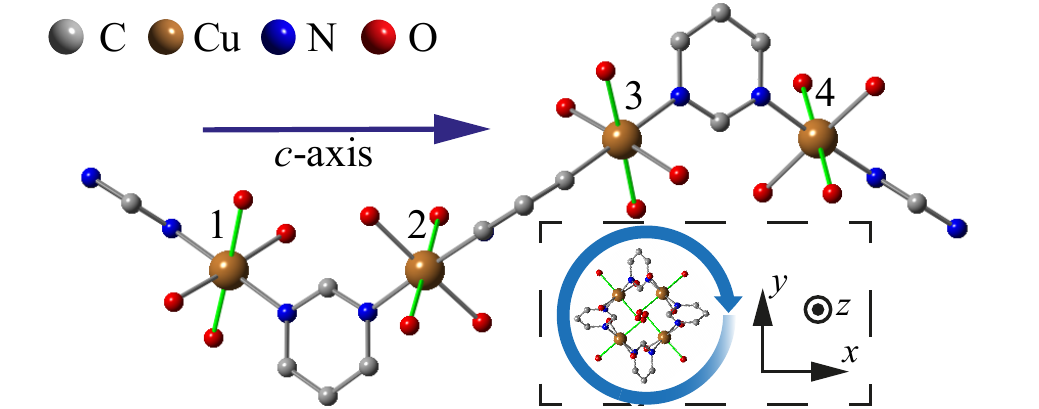}
\caption{Chain structure of [Cu(pym)(H$_2$0)$_4$]SiF$_6\cdot$H$_2$O. H atoms and [SiF$_6$]$^{2-}$ counterions omitted for clarity. The unit cell houses four inequivalent Cu(II) ions. Staggered elongated Cu\textemdash O bonds (green) correspond to the local $g_\parallel$ axes. Inset: view along $c$-axis depicting four-fold rotation of the local environment about the chain direction.}
\label{figStructure}
\end{figure}

[Cu(pym)(H$_2$O)$_4$]SiF$_6\cdot$H$_2$O crystallizes in chiral space group $P4_12_12$~\cite{CordesCrysGrowDesign2007}. Cu(II) ions are linked in chains by N\textemdash C\textemdash N moieties of pyrimidine, which propagate the intrachain exchange interactions (Fig.\,\ref{figStructure}). The Cu coordination is a distorted octahedron consisting of CuN$_2$O$_2$ equatorial planes [Cu---N 2.009(4) and 2.000(4)~\AA, Cu---O 2.043(3) and 2.091(3)~\AA], and two O atoms in the axial positions [Cu---O 2.197(3) and 2.227(4)~\AA]~\cite{CordesCrysGrowDesign2007} consistent with a Jahn-Teller elongation along this direction~\cite{Deeth86}.
In this local symmetry, $g_\parallel > g_\perp$, where $g_\parallel$ is the $g$ factor along the axial direction.
Local environments of nearest-neighbor Cu ions within a chain, and hence their local $g$ tensors, are related by a 90$^\circ$ rotation about the $c$-axis. Interchain sites are occupied by H$_2$O and SiF$_6^{2-}$ anions, leading to a minimum interchain Cu\textemdash Cu distance of 7.480~\AA.

\begin{figure*}
\begin{center}
\includegraphics[width=\textwidth]{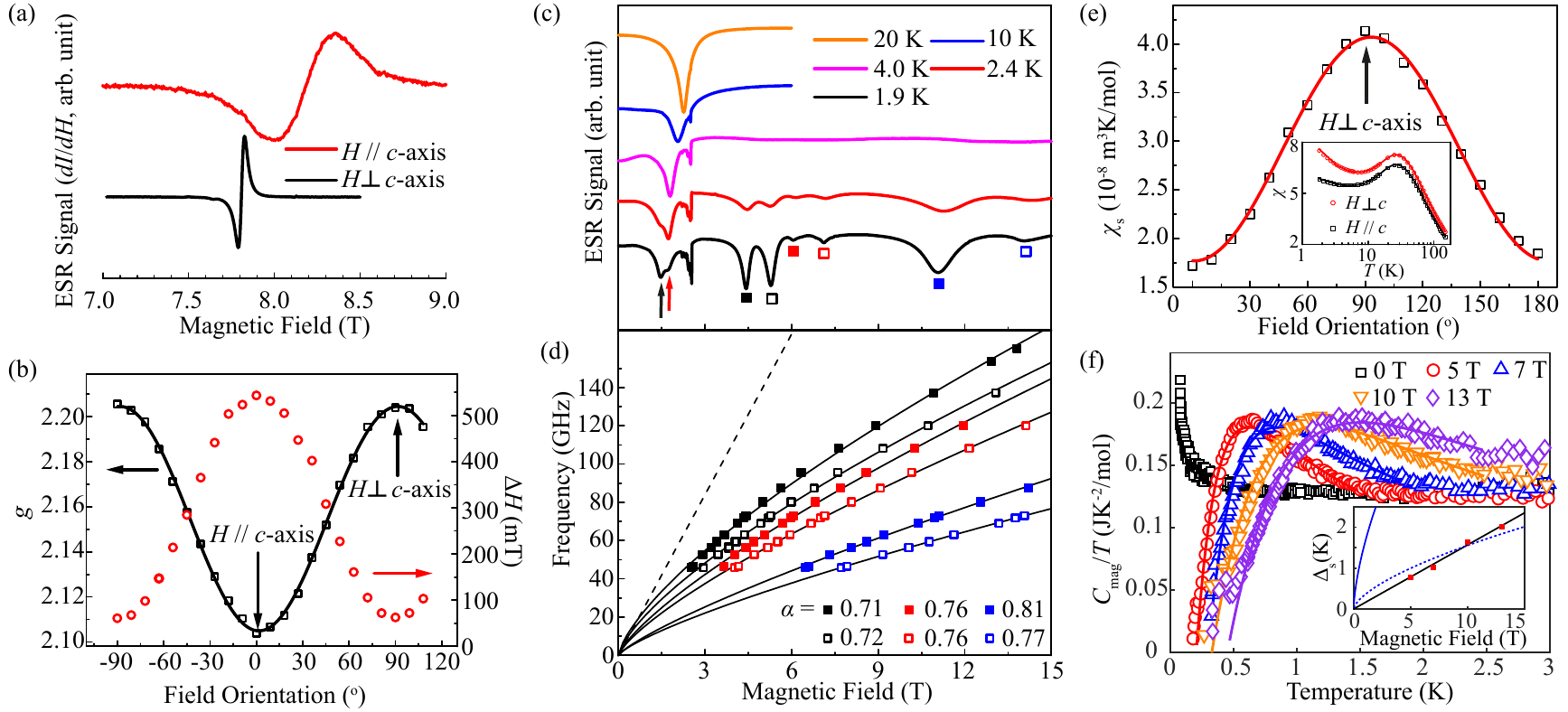}
\caption{(a) Room temperature ESR spectra at 240~GHz. A DM interaction and spin-diffusion could contribute to linewidth anisotropy~\cite{SI}. (b) Orientation dependence of Cu(II) $g$ factor ($\Box$) and linewidth ($\circ$) at 300~K. The line is a fit to $g(\theta) = (g_\mathrm{min}^2\cos^2\theta+g_\mathrm{max}^2\sin^2\theta)^{1/2}$. (c) and (d) ESR data with field $\perp c$.  (c) Temperature dependence of ESR spectra at 73~GHz. 
(d) Frequency versus field ($H$) plot showing ESR peaks observed at 1.9~K using the labeling scheme in (c). Solid lines are fits to $f = AH^\alpha$ with $\alpha$ values shown. Dashed line is the paramagnetic resonance with $g = 2$. (e) Inset: representative $\chi$ vs. $T$ data with $H = 0.2$~T. Solid line is the model $\chi(T) = \chi_\textrm{1D}(T) + \chi_\mathrm{s}/T$ described in the text. Main panel: angular variation of $\chi_s$. The line is a fit to a cos$^2$ dependence. 
(f) Temperature dependence of the magnetic contribution to heat capacity at different $H \perp c$. Lines are fits to a gapped model. Inset shows resulting field dependence of the gap (squares, size corresponds to largest $y$-axis error), a linear fit to the data (black line), the best fit to SG model (blue dotted line), and gap size predicted by SG model from experimental $g_{\rm 2s}$ value (solid blue line).}
\label{figData}
\end{center}
\end{figure*}

Room temperature ESR measurements were performed at 240~GHz employing a quasi-optic setup to probe the $g$ anisotropy. Representative spectra are shown in Fig.\,\ref{figData}(a).  A single resonance due to the strong interaction between Cu(II) ions is observed~\cite{AndersonJPSJ1954,FeyerhermJPhysCondMatt2000}. 
The measured $g$ factor is an average of the individual $g$ tensors of the four magnetically inequivalent Cu(II) sites. The strong coupling between spins and the four-fold rotational symmetry ensure an isotropic ESR angle-dependence in the $ab$-plane, hence only two principal values, $g_{\rm max} = 2.21$ and $g_{\rm min} = 2.10$, can be extrapolated from ESR data~[Fig.\,\ref{figData}(b)] and to obtain the $g$ values of the individual Cu(II) ions we must assume tetragonal local symmetry. Strictly the equatorial ligand pairs break this symmetry, but in other materials with similar Cu(II) environments for which the full $g$-tensor can be determined, it was found that $\Delta g_\perp = g_x - g_y$ is an order of magnitude smaller than $\Delta g = g_\parallel - g_\perp$~\cite{manson2008}, mitigating the tetragonal approximation. We note that relaxing this assumption would have only a small effect on the size of the staggered fields discussed later. The principal $g$ values of the individual Cu(II) ions are thus found to be $g_\parallel = 2.33$ and $g_\perp = 2.09$. These values are similar to those reported for Cu(II) ions in closely related local environments~\cite{FeyerhermJPhysCondMatt2000,manson2008}. In the $xyz$ laboratory frame~\cite{SI} the $g$ tensor of the $i$th Cu(II) ion in a unit cell can be separated into three components:
\begin{equation}
\label{Eqg}
\begin{aligned}
g_{i} &= g_{\rm u} + g_{\rm 2s} + g_{\rm 4s}\\
&= \begin{pmatrix}
2.21 & 0 & 0 \\
0 & 2.21 & 0 \\
0 & 0 & 2.10 
\end{pmatrix}
+ 0.12 \begin{pmatrix}
0 & (-1)^i & 0 \\
(-1)^i & 0 & 0 \\
0 & 0 & 0 
\end{pmatrix} \\
&+ 0.026 \begin{pmatrix}
0 & 0 & (-1)^i\delta_i \\
0 & 0 & \delta_i \\
(-1)^i\delta_i & \delta_i & 0 
\end{pmatrix}
\end{aligned}
\end{equation}
where $\delta_i =$ -1, +1, +1 and -1 respectively for $i =$ 1, 2, 3 and 4 (see Fig.~\ref{figStructure}). $g_{\rm u}$ corresponds to the uniform part of the $g$-tensor, while $g_{\rm 2s}$ is a small staggered component which repeats after every two nearest neighbour Cu(II) ions. $g_{\rm 4s}$, which we believe is so far unique to this material, is a much weaker staggered part with a four-fold period along the chain, i.e. the same periodicity as the unit cell. $g_{\rm 2s}$ is similar in form and size to that seen in Cu-benzoate and [pym-Cu(NO$_3$)$_2$(H$_2$O)$_2$]~\cite{OshikawaPRL1997a,AffleckPRB1999,FeyerhermJPhysCondMatt2000}. When a field $H$ is applied perpendicular to the chain, $g_{\rm 2s}$ will generate a staggered field that can open an excitation gap in the low-temperature spectrum. 

Further variable frequency/temperature ESR measurements were performed in a broadband spectrometer. The temperature dependence of the spectra below 20~K are shown in Fig.~\ref{figData}(c). Upon cooling, the resonance shifts to lower field and eventually splits into two as indicated by arrows. 
Below 3~K, a series of resonances (marked by squares) quickly emerge at high field (above $g = 2$) with intensities that increase rapidly with lowering temperature, confirming the ground state nature of these excitations. The temperature and frequency [Fig.~2(d)] evolution of the lines cannot be explained by paramagnetic resonances of transition metals~\cite{KrzystekCoordChemRev2006} or conventional antiferromagnetic resonances. In fact, the data are reminiscent of excitations observed in [pym-Cu(NO$_3$)$_2$(H$_2$O)$_2$], where three branches were identified as breather modes of the SG model, along with six other modes more difficult to classify~\cite{ZvyaginPRL2004}. However, we find that the frequency-field dependence of our resonances \textit{cannot} be modelled by the breather gaps proposed for SG chains~\cite{EsslerPRB2003,ZvyaginPRL2004,SI}.

Another important predicted feature due to staggered $g$-tensors is the existence of a low-temperature staggered susceptibility. Representative single-crystal susceptibility ($\chi$) measurements performed down to 1.8~K are shown in the inset to Fig.\,\ref{figData}(e). Above 10~K, the susceptibility can be well modelled as a uniform Heisenberg $S = 1/2$ chain~\cite{JohnstonPRB2000} with AF exchange $J = 42.3\pm 0.8~\textrm{K}$. This energy is consistent with the saturation field ($\approx 65$~T) observed in pulsed magnetic fields~\cite{SI}. 
The value of $J$ and lack of long-range order above 20~mK in zero field imply an upper limit of 7~mK on the size of the interchain coupling~\cite{Yasuda05}.
Below 10~K, an upturn in $\chi(T)$ emerges which can be described by an additional contribution proportional to $1/T$ and the entire temperature range can be fitted to $\chi(T) = \chi_\textrm{1D}(T) + \chi_\mathrm{s}/T$ where $\chi_\textrm{1D}(T)$ is the Heisenberg chain contribution. The staggered susceptibility $\chi_\mathrm{s}$ exhibits a pronounced angular dependence with its maximum value occurring when $H$ is applied perpendicular to the chain. The strong anisotropy of $\chi_\mathrm{s}$ confirms it is not due to paramagnetic impurities. Similar variation of $\chi_\mathrm{s}$ is observed in Cu-benzoate~\cite{Date1970} and [pym-Cu(NO$_3$)$_2$(H$_2$O)$_2$] ~\cite{FeyerhermJPhysCondMatt2000} and identified as an intrinsic property related to the staggered field~\cite{GlockePRB2006}. 

Fig.~\ref{figData}(f) shows the magnetic contribution ($C_\textrm{mag}$) to heat capacity for a deuterated sample of [Cu(pym)(H$_2$O)$_4$]SiF$_6\cdot$H$_2$O down to 100~mK. A $C\sim T^{-2}$ low-temperature tail is subtracted from the data (except the zero-field data) to account for the nuclear contribution~\cite{SI}. In zero field above 0.5~K, the nearly constant value of $C_\textrm{mag}/T$ can be interpreted as the heat capacity of a uniform $S = 1/2$ AF Heisenberg chain in the TLL state, where $C_\mathrm{mag} = 2RT/3J$~\cite{Affleck1986,Xiang1998}, giving $J = 41.9\pm1.5~\mathrm{K}$ in excellent agreement with the magnetometry data. 

The field-induced gap in [Cu(pym)(H$_2$O)$_4$]SiF$_6\cdot$H$_2$O is revealed by heat capacity measurements performed with $H\perp c$. On warming from low temperatures, $C_\textrm{mag}$ is first suppressed below, and then rises above, the zero-field curve before meeting it at high temperature. This behaviour is identical to that of the non-chiral staggered chains~\cite{DenderPRL1997,FeyerhermJPhysCondMatt2000} and indicates the emergence of an excitation gap that increases with applied field. Given the similarities with the non-chiral staggered chains, the expression for the temperature dependence of $C_{\rm mag}$ derived from the SG model will provide the best possible estimate of the gap in our system at a particular magnetic field. However, fitting our data to the SG model~\cite{essler99} yields a gap of $\Delta_{\rm s} = 1.98$\,K at 13\,T, significantly smaller than the expected value of 8.24\,K, calculated using $g_{\rm 2s}$ and $J$ values obtained from ESR and magnetometry. More importantly, the field evolution of the gap exhibits $\Delta_{\rm s} \propto H$ [inset to Fig.\,\ref{figData}(f)], which is distinctly different from the expectation of the SG model, where $\Delta_{\rm s} \propto H^{2/3}$~\cite{OshikawaPRL1997a}. 

It is clear that [Cu(pym)(H$_2$O)$_4$]SiF$_6\cdot$H$_2$O exhibits the staggered $g$-tensors, ground-state excitations and field-induced spin gap similar to those seen in the SG chains. However, the size of the gap and its linear field dependence are inconsistent with theoretical predictions for a traditional staggered system. We believe that we can qualitatively account for the departures from the existing theories based on differences between the chiral and non-chiral staggered structures.

\begin{figure}[t]
\centering
\includegraphics[width=\columnwidth]{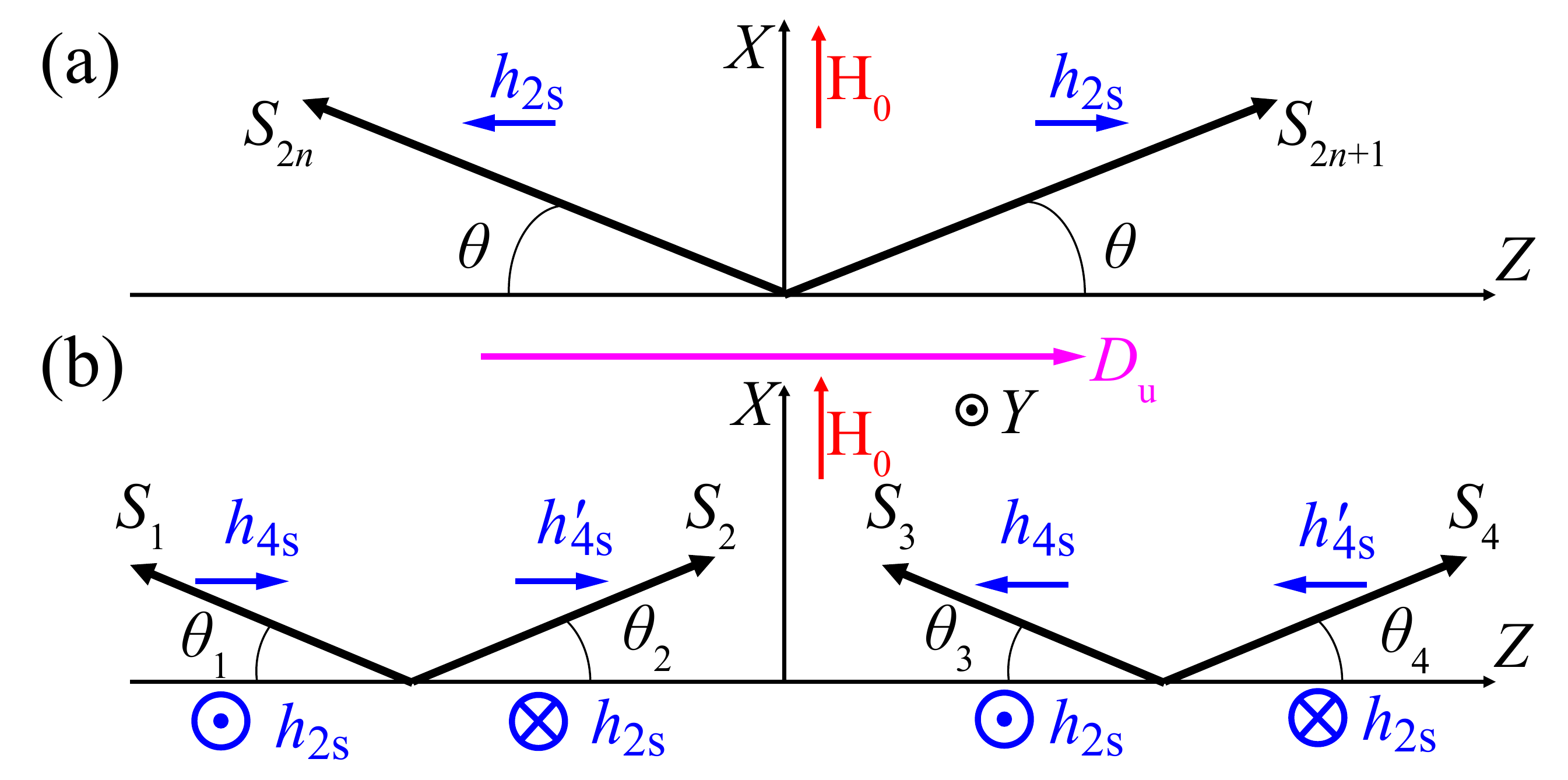}
\caption{Classical spin configurations for the unit cell of (a) Cu-benzoate and (b) [Cu(pym)(H$_2$O)$_4$]SiF$_6\cdot$H$_2$O in applied field $H_0 \parallel X$-axis. $Z$ is the chain direction. Also shown are relative directions of the two-fold and four-fold staggered fields ($h_{\rm 2s}$, $h_{\rm 4s}$ and $h^\prime_{\rm 4s}$), and uniform DM interaction ($D_{\rm u}$) allowed in [Cu(pym)(H$_2$O)$_4$]SiF$_6\cdot$H$_2$O. The $XZ$ canting in (b) would minimize coupling between the spins and $h_{\rm 2s}$.}
\label{figSpinStruc}
\end{figure}

The staggered fields in the existing model of Cu-benzoate are compared to our chiral chain in Fig.\,\ref{figSpinStruc}. In both cases the chains lie along $\textbf{Z}$ and the applied field $\textbf{H}_0\mathbin{\parallel} \textbf{X}$. For Cu-benzoate, a canted spin configuration, caused by competition between AF exchange and $H_0$, is stabilized in the $XZ$-plane by the same two-fold periodic staggered fields $\textbf{h}_{\rm 2s}\mathbin{\parallel} \textbf{Z}$ that are responsible for the gap~\cite{AffleckPRB1999}. In contrast in the chiral system, the measured $g$-tensor (Eq.~\ref{Eqg}) gives rise to two-fold staggered fields perpendicular to both $\textbf{H}_0$ and $\textbf{Z}$. Thus if the spins adopt a canting in the $XZ$ plane, the coupling to $\textbf{h}_{\rm 2s}\mathbin{\parallel} \textbf{Y}$ will be lost. This would be a simple explanation for the observed suppression of the field-induced gap, but requires an additional interaction, not present in previously studied staggered chains, that would favor $XZ$ over $XY$ canting. 
The crystal symmetry of [Cu(pym)(H$_2$O)$_4$]SiF$_6\cdot$H$_2$O permits such an interaction in the form of an equal and opposite DM interaction on nearest-neighbor chains. By symmetry, the DM interaction of a given chain can be decomposed into a uniform component parallel to the chain axis of magnitude $D_{\rm u}$, and a four-fold periodic staggered component perpendicular to the chain axis. 
Including the uniform DM term and applying $\mathbf{H}_0\parallel\mathbf{X}$, the Hamiltonian can be written as,
\begin{equation}
\label{uniformDM}
\begin{aligned}
\hat{\cal{H}} & = \sum_i J(\hat{S}_i^X\hat{S}_{i+1}^X + \hat{S}_i^Y\hat{S}_{i+1}^Y + \lambda \hat{S}_i^Z\hat{S}_{i+1}^Z)\\
& +\sum_i D_\mathrm{u}(\hat{S}_i^X\hat{S}_{i+1}^Y-\hat{S}_i^Y\hat{S}_{i+1}^X) + \sum_iH_0g_u\hat{S}_i^X \\
& +\sum_i g_{\rm 2s}H_0\hat{S}_i^Y + \sum_i g_{\rm 4s}H_0 \hat{S}_i^Z,
\end{aligned}
\end{equation}
where $J$ is the intrachain AF exchange, which can now possess a small anisotropy ($\lambda\approx 1$).

The classical model of an $S = 1/2$ chain with uniform $g_\mathrm{u}$ and $D_\mathrm{u}$ in a transverse $H_0$ predicts an incommensurate-commensurate transition at $H_0 = \pi D_\mathrm{u}S$, where the ground state at high fields is an AF canted structure lying in the $XY$ or $XZ$ plane for $\lambda < 1$ or $\lambda > 1$, respectively. Recent quantum analyses of the same model shows that $XZ$ canting is preferred for a substantial field range, even with a weak easy-plane type exchange anisotropy ($\lambda < 1$)~\cite{Garate2010,Chan2017}. In our case this would effectively mitigate the effect of the small two-fold staggered field along $Y$. At low magnetic fields, the ground state of the model is an incommensurate helical $XY$ antiferromagnet with propagation vector along $\textbf{Z}$~\cite{Aristov2000,Garate2010,Chan2017}. Averaged across the chain, this structure will also negate the effect of the $h_{\rm 2s}$ term, which is commensurate, resulting in a reduced gap. Furthermore, the quantum simulations show that a tiny field component parallel to the chain, e.g. caused by experimental sample misalignment, will stabilize the gapless $XY$ helical phase against a strong transverse field~\cite{Garate2010}, allowing the $XY$ canted phase that couples strongly to $h_{\rm 2s}$ to appear only in the presence of a considerable easy-plane anisotropy. Hence, we expect the ground state to show either an $XY$ helix or $XZ$ canting (or a mixture). In either case, the coupling to $h_{\rm 2s}$ will be significantly weakened compared to the staggered chains considered previously, leading to suppression of the field-induced gap. 

An interesting consequence of the chiral crystal structure with $XZ$ spin-canting is that while $\mathbf{h}_{\rm 2s}$ is perpendicular to the spins, the weaker staggered fields $\mathbf{h}_{\rm 4s} = g_{\rm 4s}\mathbf{H}_0$ are parallel to $\mathbf{Z}$ [Fig.~\ref{figSpinStruc}(b)] and so contribute to the excitation spectrum potentially producing a field-induced gap. To explore this, we consider a spin-wave model of an $S = 1/2$ Heisenberg chain in the $XZ$ canted configuration with mutually perpendicular uniform and staggered fields. As noted earlier, in principle $g_{x} \neq g_{y}$, meaning that $h_{\rm 4s}$ acting on neighbouring ions need not be identical. Nevertheless, the space group symmetry of [Cu(pym)(H$_2$O)$_4$]SiF$_6\cdot$H$_2$O requires that: (i) these staggered fields have a four-fold repetition; and (ii) next-nearest neighbors within a chain, which are related by a $180^\circ$ rotation, must experience equal, but opposite values. This is different to the staggered fields that open the gap in Cu-benzoate, where $\mathbf{h}_{\rm 2s}$ takes opposite values on adjacent sites [Fig.~\ref{figSpinStruc}(a)]. In both cases, the exchange interaction ensures that the magnetic moment along the chain is antiparallel for neighboring ions. The first order spin-wave approximation suggests that the spin-canting angle only depends on the external field $H_0\gg h_{\rm 4s}$ and the interaction $J$, specifically $\theta_i = \theta = \sin^{-1} (H_0/4JS)$ for all spins. Thus the classical energy contribution of the four-fold staggered field is: 
\begin{equation}
\label{EqEnergy1}
E_\mathrm{\rm 4s} = \cos{\theta}(h_{\rm 4s}S - h_{\rm 4s}^\prime S - h_{\rm 4s}S + h_{\rm 4s}^\prime S) = 0,
\end{equation} 
where $h_{\rm 4s}$ and $h_{\rm 4s}^\prime$ are the staggered-field values on odd and even sites, respectively. This complete cancellation implies that to first order in the spin-wave approximation, the four-fold staggered field in [Cu(pym)(H$_2$O)$_4$]SiF$_6\cdot$H$_2$O has no contribution to its energy spectrum, whereas in Cu-benzoate the field-induced gap is readily seen in the first-order perturbation~\cite{OshikawaPRL1997a}. Only when the analysis is extended to second order by considering the corrections to the canting angles $\theta_j$ due to $\textbf{h}_{\rm 4s}$ and $\textbf{h}_{\rm 4s}^\prime$ will a gap start to emerge~\cite{SI}. 

Summarizing, we propose the key differences between the nature of the gap observed in chiral [Cu(pym)(H$_2$O)$_4$]SiF$_6\cdot$H$_2$O and the traditional staggered chains reside in a low-temperature spin configuration that decouples the system, fully or partially, from the two-fold staggered field, and is favored by the presence of a uniform DM interaction. In the event of a mixture of $XY$, $XZ$ or helical configurations, the effect of $h_{\rm 2s}$ on the excitation spectrum is still important, but will be altered from existing models. In addition, and particularly for the situation of complete decoupling of the spins from $h_{\rm 2s}$, the added presence of the four-fold staggered field necessitates new theoretical predictions for the size and field-dependence of the gap, as well as for the excitation modes seen in ESR. We note it is possible that in applied field the $XZ$ canted phase itself could exhibit a gap in the absence of staggered fields~\cite{suhas2008}. 

The purely $XZ$ canting shown in Fig.\,\ref{figSpinStruc}(b) relies on the uniform DM interaction being significantly stronger than $h_{\rm 2s}$. We expect $D_{\rm u} \propto (\Delta g/g)J \gg h_{\rm 2s}$ for the experimental field range, and the smooth linear evolution of the gap up to 13~T [inset to Fig.~\ref{figData}(f)] suggests no abrupt spin orientation takes place, however the data do not permit an unambiguous estimate of $D_{\rm u}$. We point out again that a staggered DM interaction is allowed by the symmetry of [Cu(pym)(H$_2$O)$_4$]SiF$_6\cdot$H$_2$O and must be considered in any future theoretical model. 

Our results demonstrate that spin chains in which local magnetic environments are related via screw symmetry can present a remarkable suppression of the field-induced SG spin gap via emergence of uniform DM interactions and complex staggered fields. This opens up the possibility of searching for other materials where anisotropic interactions and particular crystal symmetries conspire to enable entirely novel magnetic states.

We thank EPSRC for support. PAG acknowledges that this project has received funding from the European Research Council (ERC) under the European Union's Horizon 2020 research and innovation programme (grant agreement No. 681260). RDJ acknowledges financial support from the Royal Society. A portion of this work was performed at the National High Magnetic Field Laboratory, which is supported by National Science Foundation Cooperative Agreement No. DMR-1157490 and the State of Florida, as well as the  {\it Strongly Correlated Magnets} thrust of the DoE BES ``Science in 100 T'' program. The work at EWU was supported by the NSF through grant no. DMR-1703003. We are grateful for the provision of beam time for muon measurements at the STFC ISIS Facility, UK and the Swiss Muon Source, Paul Scherrer Institut, Switzerland, and to F.L. Pratt and C. Baines for experimental assistance. Roberts Williams is thanked for useful discussions. Data presented in this paper resulting from the UK effort will be made available at XXXXXXX.

\newpage
~
\newpage

%%%%%%%%%%%%%%%%%%%%%%%%%%%%%%%%%%%%%%%%%%%%%%%%%%%%%%

\onecolumngrid

\setcounter{equation}{0}
\setcounter{figure}{0}
\setcounter{table}{0}
\renewcommand{\theequation}{S\arabic{equation}}
\renewcommand{\thefigure}{S\arabic{figure}}
\renewcommand{\thetable}{S\Roman{table}}
\newcommand{\CuII}{Cu$^{2+}$}

\renewcommand{\thesubsection}{\thesection.\alph{subsection}}

\begin{center}
{\Large \sc \bf Supplemental Material}
\end{center}

\section{Samples}
Crystals of [Cu(pym)(H$_2$O)$_4$][SiF$_6$]H$_2$O grow as rod-like blocks with the long axis of the crystal parallel to the Cu---pym chain. For the heat capacity measurements a sample of dimensions $7 \times 2 \times 1 $~mm$^3$ was used. For susceptibility measurements the sample was $6 \times 1.5 \times 1.5$~mm$^3$. The ESR measurements made use of crystals with similar aspect ratios. For the room-temperature measurements the sample length was 1.5~mm and for the low-temperature ESR data the sample was 2~mm long.

\section{Muon-spin relaxation}

Zero field (ZF) muon-spin relaxation ($\mu^{+}$SR) measurements were made at the ISIS Facility, Rutherford Appleton Laboratory, UK using the EMU instrument and at the Swiss Muon Source, Paul Scherrer Institut, CH using the LTF instrument. The polycrystalline sample was mounted on a silver plate using a thin layer of silicon grease. In a $\mu^{+}$SR experiment \cite{Blundell1999} spin-polarized positive muons are stopped in a target sample, where the muon usually occupies an interstitial position in the crystal. The observed property in the experiment is the time evolution of the muon spin polarization, the behaviour of which depends on the local magnetic field $B$ at
the muon site. Each muon decays, with a lifetime of 2.2~$\mu$s, into two neutrinos and a positron, the latter particle being emitted preferentially along the instantaneous direction of the muon spin. Recording the time dependence of the positron emission directions
therefore allows the determination of the spin-polarization of the ensemble of muons. In our experiments positrons are detected by detectors placed forward (F) and backward (B) of the initial muon polarization direction. Histograms $N_{\mathrm{F}}(t)$ and $N_{\mathrm{B}}(t)$ record the number of positrons detected in the two detectors as a function of time $t$ following the muon implantation. The quantity of interest is the decay positron asymmetry function, defined as
\begin{equation}
\label{usramp}
A(t)=\frac{N_{\mathrm{F}}(t)-\alpha_{\mathrm{exp}} N_{\mathrm{B}}(t)}
{N_{\mathrm{F}}(t)+\alpha_{\mathrm{exp}} N_{\mathrm{B}}(t)} \, ,
\end{equation}
where $\alpha_{\mathrm{exp}}$ is an experimental calibration constant. $A(t)$ is proportional to the spin polarization of the muon ensemble.

\begin{figure}[h]
\begin{center}
\includegraphics[width=0.4\columnwidth]{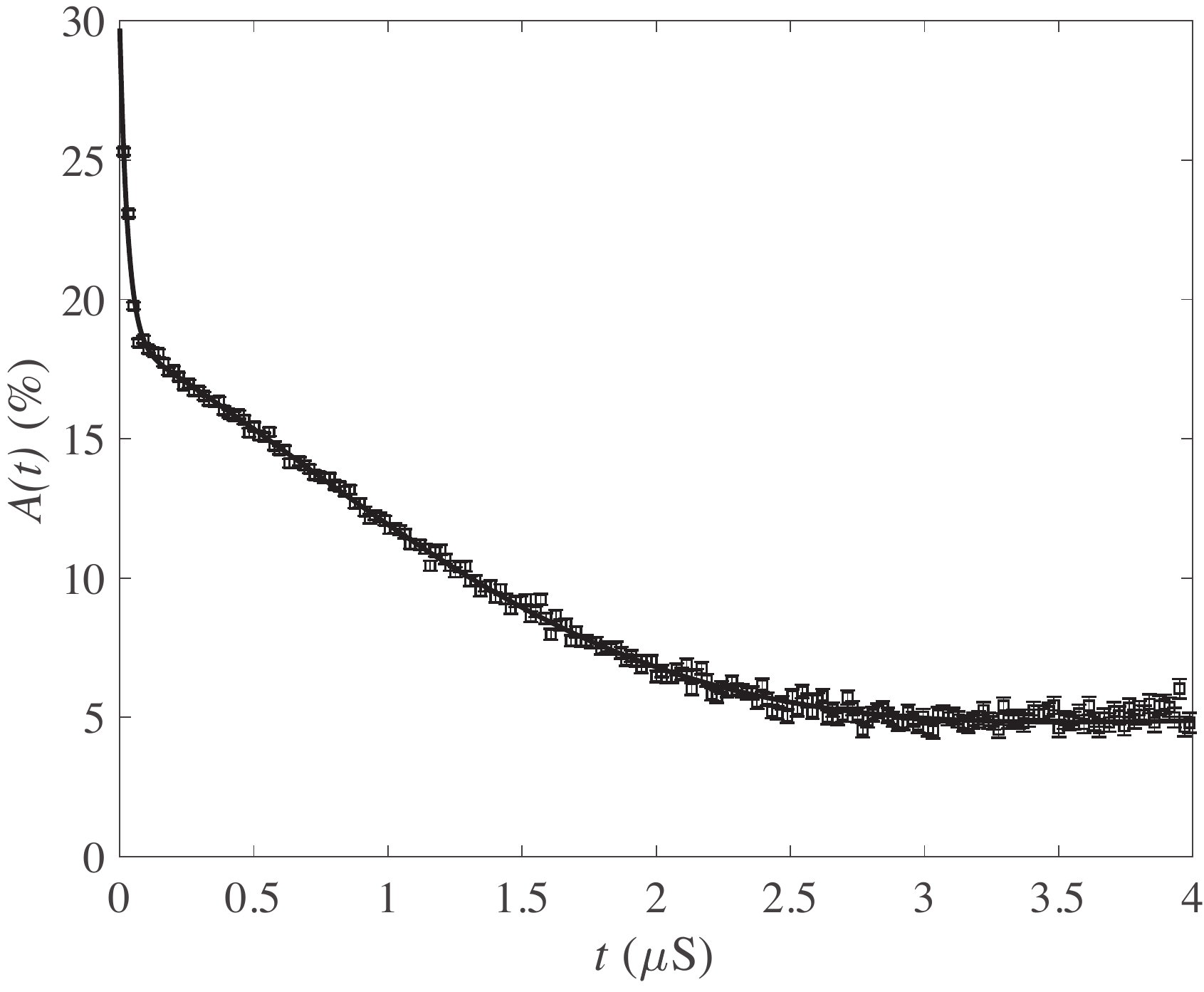}
\caption{(a) ZF $\mu^{+}$SR spectrum measured at 20~mK showing a contribution from two components. A fit is shown to Eq.~ (\ref{usrfit}).  
\label{muonfig1}}
\end{center}
\end{figure}

The ZF spectrum measured at 20~mK is shown in Fig.~ \ref{muonfig1}. We note that no oscillations or related signatures are observed in the muon asymmetry, which strongly suggests that the material does not magnetically order at temperatures above 20~mK. The measured spectra are found to contain two contributions. The first is a fast relaxing component $A_{1}$ which dominates the signal at early times and is well described by an exponential function $\exp(-\lambda t)$. The second is a larger, slowly relaxing component $A_{2}$, which dominates at intermediate times and fits to the Kubo-Toyabe (KT) function $f_{\mathrm{KT}}(\Delta, B_{\mathrm{app}}, t)$ \cite{Hayano1979} where $\Delta$ is the second moment of the static, local magnetic field distribution defined by $\Delta = \gamma_{\mu} 
\sqrt{\langle (B - \langle B \rangle)^{2} \rangle}$, $B_{\mathrm{app}}$ is the magnitude of the applied longitudinal magnetic field and $\gamma_{\mu}$($=2 \pi \times 135.5$~MHz T$^{-1}$) is the muon gyromagnetic ratio. The KT function is characteristic of spin relaxation
due to a random, quasi-static distribution of local magnetic fields at diamagnetic muon sites. We do not observe the recovery in asymmetry at late times that is expected for the static KT function. The lack of this recovery is probably due to slow dynamics in the random field distribution and is crudely modelled here with an exponential term $\exp(-\Lambda t)$. The data were found to be best fitted with the resulting function over the entire time range
\begin{equation}
\label{usrfit}
A(t)=A_{1}f_{\mathrm{KT}}(\Delta, B_{\mathrm{app}}, t)\exp(-\Lambda t) 
+ A_{2} \exp(-\lambda t)
+A_{\mathrm{bg}}
\end{equation}
where $A_{\mathrm{bg}}$ represents a constant background signal  from those muons that stop in the sample holder. The fitted magnitude of $\Delta \approx 0.5$~MHz suggests that the random magnetic field distribution giving rise to the KT function is due to nuclear magnetic moments, implying that the field due to electronic moments at these muon sites is motionally narrowed out of the spectrum due to very rapid fluctuations. The observation of two distinct components, $A_{1}$ and $A_{2}$, in the asymmetry spectrum suggests that there are two distinct muon stopping states in the material. We note that behaviour consistent with two muon sites has been observed previously in copper chain compounds with similar structures.

From the observation that the material does not undergo a magnetic ordering transition above 20~mK, it is possible to put an upper limit on any interchain interaction present in the system. The onset temperature of long-range order in highly one-dimensional $S=1/2$ antiferromagnets depends critically on the ratio of the interchain ($J_\perp$) to intrachain ($J$) exchange strengths~\cite{Yasuda05}. Using the relation given in that reference, the lack of long-range order indicates that $J_\perp/J < 1.7\times10^{-4}$ or $J_\perp < 7$~mK.

\section{Crystal symmetry and Dzyaloshinskii-Moriya interactions}

As for all exchange interactions present in the spin Hamiltonian, in the absence of symmetry-breaking order or external fields, the antisymmetric Dzyaloshinskii-Moriya (DM) interaction must respect the symmetry of the parent crystal structure. The relevant term in the mean-field energy is typically written $\mathbf{D}\cdot(\hat{\mathbf{S}_i}\times\hat{\mathbf{S}_j})$, where the DM interaction is represented by a time-even and parity-even pseudo vector (axial vector), $\mathbf{D}$.

The crystal structure of [Cu(pym)(OH$_2$)$_4$][SiF$_6$]H$_2$O has space group $P$4$_1$2$_1$2, and is composed of two chiral Cu(pyrimidine) chains per unit cell with a common, global chirality. The two chiral chains are related by the 2$_1$ and 2 symmetry operators of the space group (note that both 2$_1$ and 2 operations do not switch structural chirality, which is formally represented by a time-even, parity-odd pseudo scalar). An individual chain has symmetry 4$_1$, such that 4 Cu atoms bonded together via pyrimidine molecules complete a full chiral rotation per unit cell. When occurring by itself, 4$_1$ symmetry supports a uniform $\mathbf{D}_\mathrm{u}$ parallel to the axis of the chain. The 4$_1$ symmetry also supports a four-fold staggered $\mathbf{D}_\mathrm{s}$ lying perpendicular to the uniform $\mathbf{D}_\mathrm{u}$, which transforms in the same way as the Cu-(pyrimidine)-Cu bonds. 

Without loss of generality, the DM interaction in [Cu(pym)(OH$_2$)$_4$][SiF$_6$]H$_2$O can be decomposed into the two orthogonal components, uniform and staggered, described above. We note, however, that the two terms in the mean-field energy will have common components of the spins, such that they cannot be minimised independently.

Finally, we note that space group $P$4$_1$2$_1$2 does not support a net $\mathbf{D}$. The 2$_1$ and 2 symmetry operators give the same configuration of staggered $\mathbf{D}_\mathrm{s}$ on every chain, but opposite signs of the uniform  $\mathbf{D}_\mathrm{u}$ for the two chiral chains in each unit cell.

\section{Electron spin resonance} 

\subsection{Room-temperature electron spin resonance and staggered $g$ tensors}

\begin{figure}[t]
	\centering
	\includegraphics[width=0.4\columnwidth]{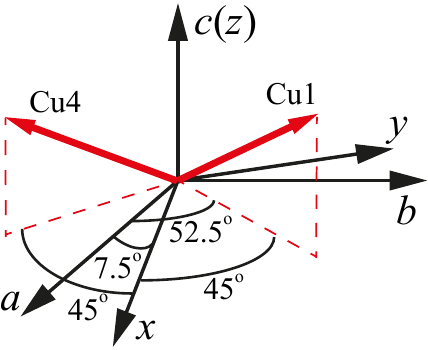}
	\caption{Schematic plot showing the relationship between the crystalline $abc$ axes and the laboratory frame $xyz$. The red arrows represents the JT axes of Cu1 and Cu4, i.e. the local principal $g$ axes, and the dashed lines indicate their projection into the $ab$ plane. The projection of the JL axis for Cu1 is $52.5^\circ$ away from the $a$ axis. The $x$ axis lies halfway between the  projections of the JT axes for Cu1 and Ch4, making it $7.5^\circ$ away from the $a$ axis.}
	\label{coordinate}
\end{figure}

In order to separate the uniform and staggered parts of the $g$ tensors, it is convenient to define a laboratory frame $xyz$ based on the Jahn-Teller (JT) axes of the Cu(II) atoms. This $xyz$ coordination is chosen such that the $g$ tensors for each \CuII spins can be separated into a uniform diagonal part, invariant for all spins, and staggered off-diagonal parts. To achieve this, the $x$ axis is set to be midway between the projections of two adjacent JL axes in the $ab$ plane. Therefore, the $x$ axis lies within the $ab$-plane and it is rotated $7.5^\circ$ away from the $a$ axis. The $z$ axis is parallel to the crystallographic $c$ axis.  The conversion between the crystalline $abc$ axes and the $xyz$ frame is shown in Fig.~\ref{coordinate} .

\vspace{10pt}

An anisotropic $g$ tensor is expected for the Cu(II) spins due to the JT distorted octahedral coordination environment surrounding the Cu(II) ions. The JT axis, defined as the elongated Cu---O bond, lies $9.1^\circ$ out of the $xy$ plane. The local principal axis of the $g$ tensor is expected to be parallel to the JT axis. Assuming tetragonal local symmetry, the local $g$ tensor is expected to take the diagonalized form:
\begin{equation}
\label{g_local}
\begin{aligned}
g_\mathrm{loc} &= \begin{pmatrix}
g_\perp & 0 & 0 \\
0 & g_\perp & 0 \\
0 & 0 & g_\parallel  
\end{pmatrix}\\
\end{aligned}
\end{equation}
with the $g_\parallel$ component along the  elongated JT axis ($g_\parallel  > g_\perp$). The local $g$ tensors for individual spins can be transformed to the global $xyz$ coordination with the following Euler rotations:
\begin{equation}
\label{transform}
\begin{aligned}
g_i^{xyz} & = Q_i^T g_\mathrm{loc} Q_i
\end{aligned}
\end{equation}
with
\begin{equation}
\label{euler}
\begin{aligned}
Q_i &= \begin{pmatrix}
1 & 0 & 0 \\
0 & \cos{\theta} & \sin{\theta} \\
0 & -\sin{\theta} & \cos{\theta} 
\end{pmatrix}
\begin{pmatrix}
\cos{\phi_i} & \sin{\phi_i} & 0 \\
-\sin{\phi_i} & \cos{\phi_i} & 0 \\
0 & 0 & 1  
\end{pmatrix}\\
\end{aligned}
\end{equation}
$\theta = 80.9^\circ (90 - 9.1)$, which is the tilting angle between the JT axis and the $z$ axis. $\phi_ i = 45^\circ, 135^\circ, 225^\circ$ and $315^\circ$ for $i = 1, 2, 3$ and 4, respectively, in consistent with the four-fold rotation of the local coordination environment about the $z$ axis. In the strong coupling limit, the average of the four Cu(II) sites within a unit cell give the $g$ values observed in ESR measurements [Fig.~2(a) in the main text]:
\begin{equation}
\label{g_ave}
\begin{aligned}
\frac{1}{4}\sum_{i = 1}^4 g_i^{xyz} &= \begin{pmatrix}
2.21 & 0 & 0 \\
0 & 2.21 & 0 \\
0 & 0 & 2.10  
\end{pmatrix}\\
\end{aligned}
\end{equation}
By combing Eq.~\ref{g_local}-\ref{g_ave}, we obtained $g_\parallel = 2.33$ and $g_\perp = 2.09$. Therefore, according to Eq.~\ref{transform}, the $g$ tensors for individual spins in the $xyz$ coordination frame is:
\begin{equation}
\label{g_ind}
\begin{aligned}
g_1^{xyz} &= \begin{pmatrix}
2.210 & -0.12 & 0.026 \\
-0.12 & 2.21 & -0.026 \\
0.026 & -0.026 & 2.10 
\end{pmatrix}
\quad g_2^{xyz}  = \begin{pmatrix}
2.210 & 0.12 & 0.026 \\
0.12 & 2.21 & 0.026 \\
0.026 & 0.026 & 2.10
\end{pmatrix}\\
g_3^{xyz} &= \begin{pmatrix}
2.210 & -0.12 & -0.026 \\
-0.12 & 2.21 & 0.026 \\
-0.026 & 0.026 & 2.10 
\end{pmatrix}
\quad g_4^{xyz}  = \begin{pmatrix}
2.210 & 0.12 & -0.026 \\
0.12 & 2.21 & -0.026 \\
-0.026 & -0.026 & 2.10
\end{pmatrix}\\
\end{aligned}
\end{equation}
Which corresponds to Eq.~1 in the main text. When a transverse field is applied parallel to the $x$ axis, $\mathbf{H}_ 0 = [H_0, 0, 0]$, the effective fields ($\mathbf{H}_i$) on the \textit{i}th Cu site is:
\begin{equation}
\label{fields_x}
\begin{aligned}
\mathbf{H}_1 =  
\begin{pmatrix}
2.21H_0\\
-0.12H_0\\
0.026H_0
\end{pmatrix},\,
\mathbf{H}_2 =  
\begin{pmatrix}
2.21H_0\\
0.12H_0\\
0.026H_0
\end{pmatrix},\,
\mathbf{H}_3 =  
\begin{pmatrix}
2.21H_0\\
-0.12H_0\\
-0.026H_0
\end{pmatrix},\,
\mathbf{H}_4 =  
\begin{pmatrix}
2.21H_0\\
0.12H_0\\
-0.026H_0
\end{pmatrix}
\end{aligned}
\end{equation} 
Eq.~\ref{fields_x} is the field configuration in Fig.~3(b) of the main text. The two-fold staggered fields are in the $y$ direction while the four-fold staggered fields lies parallel to the $z$ axis.

\vspace{10pt}

The room-temperature ESR resonance has the Lorentzian shape expected for an exchange-coupled system. The linewidth ($\Delta H$) [Fig.~2(b) in the main text] has a strong angular dependence, probably due to the staggered $g$ tensors as well as DM interactions~\cite{Castner1971,Zorko2004,Herak2011}. In the high temperature limit ($k_{\rm B} T \gg J$), when the isotropic exchange interactions is large compared to the ESR linewidth ($J \gg g\mu_B\Delta H$), the ESR resonance is expected to have Lorentzian profile with a linewidth~\cite{Castner1971}:
\begin{equation}
\label{ESR_width}
\Delta H = \frac{2\pi}{\sqrt{6}}\frac{k_B}{g\mu_B}\left( \frac{M_2^3}{M_4}\right)^{1/2},
\end{equation}
$M_2$ and $M_4$ are the second and fourth moment of the resonance and their explicit expressions can be found in Ref.~\cite{Castner1971}. 

The wide ESR resonance suggests the existence of sizable DM interactions. However, we found it is difficult to reproduce the extreme orientation dependence of $\Delta H$, which varies between 50---550~mT. For instance, with $D_\mathrm{u} = 1.5$~K, $\Delta H = 50$~mT for $H$ perpendicular to the $c$ axis. Such a value of $D_\mathrm{u}$ is reasonable as the magnitude of the DM interaction is expected to be proportional to $J\times\delta g/g$, where $\delta g$ is the anisotropy in $g$. On the other hand, this $D_\mathrm{u}$ would lead to $\Delta H = 75$~mT for $H \parallel c$, which is significantly smaller than the experimental result. Furthermore, we found it is impossible to simulate the angular dependence solely based on Eq.~\ref{ESR_width} with any combination of $D_\mathrm{u}$ and $D_\mathrm{s}$. This suggests additional broadening effects need to be considered to explain the result. One possible candidate is spin diffusion due direct flip-flop processes~\cite{Balian2014,Shiddiq2016}. These energy-conserving processes are maximal for $H\parallel c$ such that all spins have the same resonance frequency, leading to extra broadening of the ESR signal. When the field is rotated away from the $c$-axis, the resonance frequencies of the neighboring spin centers differ, suppressing the extra broadening effect.

\subsection{Low-temperature ESR results}

\begin{figure}[t]
	\centering
	\includegraphics[width=0.8\columnwidth]{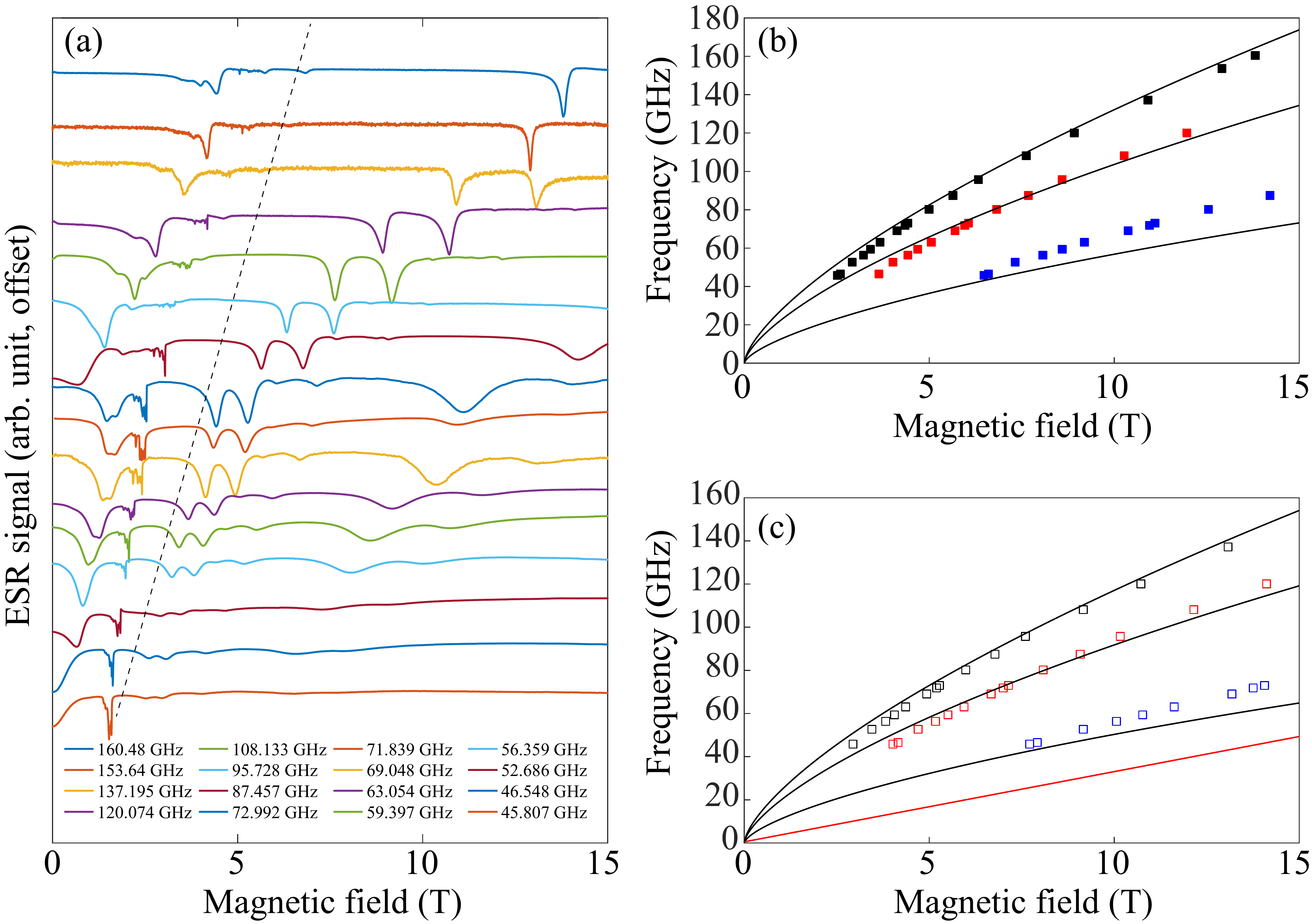}
	\caption{(a) ESR spectra for Cu(pym)(H$_2$O)$_4$]SiF$_6\cdot$H$_2$O recorded at 1.9~K with the field applied perpendicular to the crystalline $c$-axis. The spectra are separated into two regimes, the low-field part and the high field part, indicated by the dashed line in (a). (b) and (c) Frequency vs field plot showing the high-field ESR transitions following the label scheme used in Fig.~2(c) in the main text. The solid black lines are the best fit to the data using the sine-Gordon model described in Eq.~\ref{SG_ESR} and \ref{SG_soliton}. The red line shows the linear dependence of the energy gap $\Delta$  [Fig.~2(f) of the main text] as determined from heat capacity measurements and converted into GHz units. This is everywhere lower than the ESR excitation frequencies, implying that the minimum gap could be at a point in $k$-space not probed by ESR.}
\label{ESR_LT}
\end{figure}

The low-temperature multiple frequency ESR measurements were performed with a transmission type ESR spectrometer using a millimeter vector network analyzer as the microwave source and detector. The experiments were performed with the applied field perpendicular to the crystal $c$ axis. For a given frequency, the spectrum can be divided into the low-field and high-field regimes, as depicted by the dashed line in Fig.~\ref{ESR_LT}(a). In the high-field side of the spectra, six strong resonances can be resolved and followed through the experimental frequency range. As we discussed in the main text, these six resonances are reminiscent of the breather excitations observed in [pym-Cu(NO$_3$)$_2$(H$_2$O)$_2$], albeit only three breather modes were expected for the experimental field range.

One possible reason for the extra high-field resonances in our material is that the unit cell of Cu(pym)(H$_2$O)$_4$]SiF$_6\cdot$H$_2$O contains four inequivalent \CuII spins, double that for [pym-Cu(NO$_3$)$_2$(H$_2$O)$_2$], although it is not obvious exactly how the larger unit cell may lead to the additional resonances. Another possible explanation is that the breather modes are affected by a field parallel to the chain axis, e.g. due to a small field misalignment. This longitudinal field causes different effects to the chains with opposite $\mathbf{D}_\mathrm{u}$. In this case, the signal could be due to two groups of breather excitations, each with three modes as predicted by the SG model, arising due to differences in the effective staggered fields between the two subspecies with opposite $\mathbf{D}_\mathrm{u}$. This could also account for the similar amplitudes of the resonances marked by open/filled black squares [see Fig.~2(c) in main text]. 

To explore this possibility within the existing model of non-chiral chains, attempts were made to fit the high-field ESR resonances with two sets of breathers modes with different parameters. In the SG model, the $n$th breather gap is expected to be~\cite{Zvyagin2004}:
\begin{equation}
\label{SG_ESR}
\begin{aligned}
\Delta_n = 2\Delta_s\sin{n\pi\xi/2}
\end{aligned}
\end{equation}
where $\Delta_s$  is the soliton gap 
\begin{equation}
\label{SG_soliton}
\begin{aligned}
\Delta_s = J \frac{2\Gamma(\xi/2)v_F}{\sqrt{\pi}\Gamma[(1 + \xi)/2]}\left[ \frac{g\mu_BH}{Jv_F} \frac{\pi\Gamma[1/(1+\xi)]cA_x}{2\Gamma[\xi/(1+\xi)]} \right]^{(1+\xi)/2}
\end{aligned}
\end{equation}
At a given field $H$, there are $n = 1, ... [1/\xi]$ breather branches. $\xi(H/J)$, $v_F(H/J)$ and $A_x(H/J)$ in Eq.~\ref{SG_ESR} and \ref{SG_soliton} are known~\cite{Essler2003}. The value of $\xi(H/J)$ varies between 0.3065 and 0.2585 in the experimental field range; hence three branches are expected. $J$ is the antiferromagnetic interaction between spins and is set to 42.3~K (see main text). The only free parameter in the fitting is the effective staggered field coupling $c = h_\mathrm{2s}/H$, which should be close to $g_\mathrm{2s}$. 

The best fits are shown in by the solid lines in Fig.~\ref{ESR_LT}(b) and (c). We found it is impossible to simultaneously fit all three branches for either set of resonances. In addition, within each branch, the field dependence of the resonances also deviate from the SG model considerably. Furthermore, the fits gives the $c$ value of 0.011 and 0.013, significantly smaller than the measured $g_\mathrm{2s}$. Therefore, we conclude it is impossible to model the low-temperature ESR spectra with the SG model proposed for the two-fold staggered chains~\cite{Zvyagin2004,Essler2003}.

\vspace{10pt}

Analyzing the low-field transitions in Fig.~\ref{ESR_LT}(a) is challenging due to the difficulty in classifying the resonances within the experimental frequency/field range. In most spectra, a relatively broad peak is observed at low fields end with a fine structure. As shown in Fig.~2(c) in the main text, upon cooling, the paramagnetic resonance shifts towards lower field and eventually splits into two peaks. A similar low-temperature evolution of the ESR spectra has been reported in Cs$_2$CuCl$_4$ and K$_2$CuSO$_4$Br$_2$, which is attributed to the splitting of the spinon continuum around $k = 0$ when the magnetic field has a component along the DM interaction~\cite{PovarovPRL2011,Smirnov2015}. However, the four-fold staggered $\mathbf{D}_\mathrm{S}$ configuration makes it impossible for the field to be either parallel or perpendicular to all DM vectors, complicating the ESR spectra. On the other hand, it is challenging to identify the other sharp resonances observed in the low-field part of the spectra. Similar resonances were observed in the ESR study for [pym-Cu(NO$_3$)$_2$(H$_2$O)$_2$] and were speculated to be related to chain-edge effects~\cite{Zvyagin2004,Zvyagin2012}.

\section{Heat capacity}

\subsection{Subtraction of nuclear Schottky anomaly.} 

\begin{figure}[t]
	\centering
	\includegraphics[width=0.75\columnwidth]{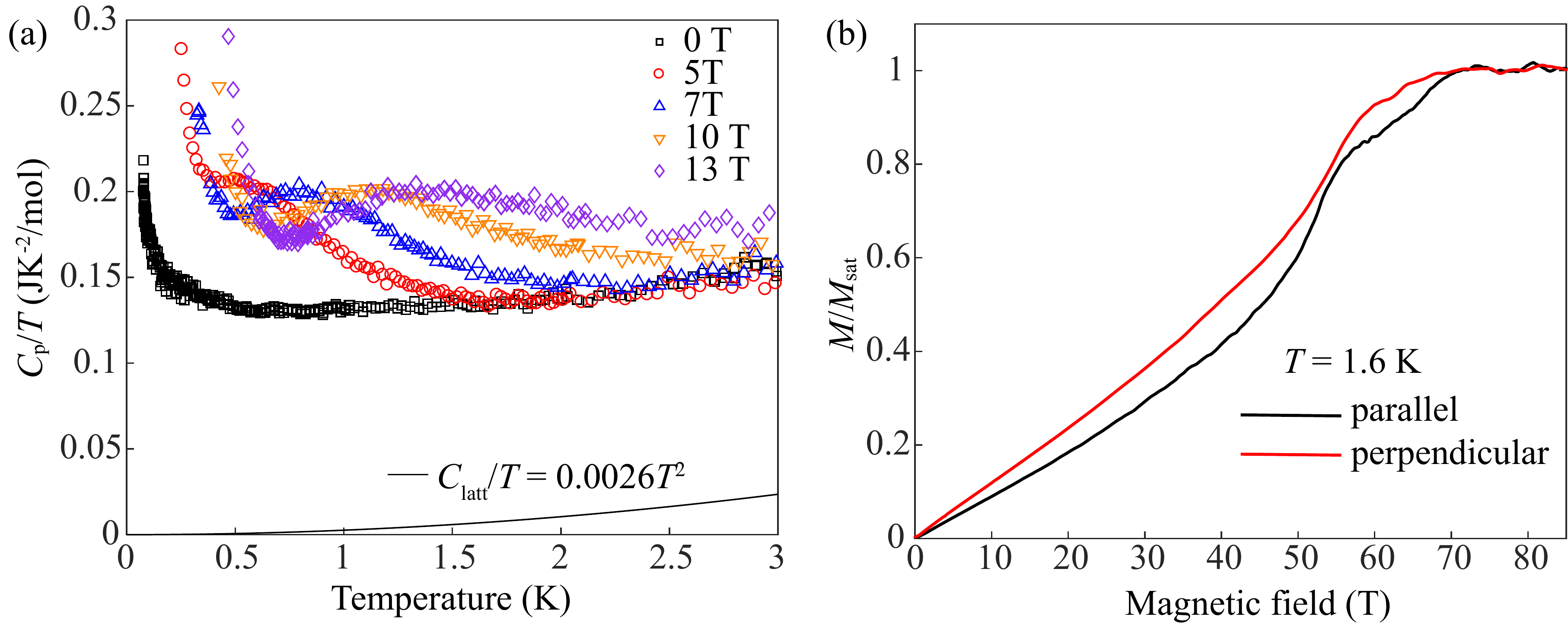}
	\caption{(a) Temperature dependence of the heat capacity, $C_\mathrm{p}$, in different transverse fields ($H \perp c$). A broad peak was observed around 0.52, 0.81, 1.1 and 1.3~K for $H = 5$, 7, 10 and 13~T, respectively. The solid line is the calculated lattice contribution. (b) Magnetization of [Cu(pym)(H$_2$O)$_4$]SiF$_6\cdot$H$_2$O divided by the saturation magnetization $M_{\rm sat}$ taken using pulsed magnetic fields at 1.6~K. The field is applied both parallel (black line) and perpendicular (red line) to the chain axis. The saturation field is consistent with the value of the intrachain exchange constant as determined from both heat capacity and low-field magnetic susceptibility.}
	\label{Cp_and_M}
\end{figure}

The heat capacity ($C_\mathrm{p}$) was measured as a function of temperature and magnetic field. In order to minimize the contribution from nuclear spins at low temperatures, the measurement was performed with a deuterated sample, [Cu(pym-D)(D$_2$O)$_4$]SiF$_6\cdot$D$_2$O. Fig.~\ref{Cp_and_M}(a) shows $C_\mathrm{p}/T$ recorded with the field applied perpendicular to the crystalline $c$ axis. At zero field, for $T > 0.8$~K, $C_\mathrm{p}$ can be well described with $C_\mathrm{p} = \alpha T + \beta T^3$, where the first and second terms correspond to the 1D spin correlation and the lattice contribution, respectively. Fitting the zero-field data gives $C_\mathrm{latt}= \beta T^3$ with $\beta = 2.6$~mJ/(mol K$^4$). This $C_\mathrm{latt}$ contribution is independent of the applied field and is removed from all data.

Upon cooling the sample, $C_\mathrm{p}$ exhibits a board peak when an external field is applied before a sharp upturn at the lowest temperatures. Similar behavior was observed in the hydrogenated sample. The broad peak is related to the field induced gap discussed in the main text.  On the other hand, We found the low temperature upturns the can be described as $C_\mathrm{p} \propto T^{-2}$ and is likely due to a nuclear Schottky effect~\cite{An2010}. The nuclear heat capacity $C_\mathrm{n}$ is expected to follow $ C_\mathrm{n} = (a_0+a_1H^2)/T^2$. $a_0$ is due to nuclear quadrupole splitting of nuclei with nuclear spins $I > 1/2$ while $a_1$ is the nuclear Zeeman energy. 

At zero field, the $\mu^+$SR data shown no evidence of magnetic order above 20~mK and no spin gap is expected for a $S = 1/2$ AFM chain. Therefore, below 0.2~K, the zero-field heat capacity data is fitted with $C_\mathrm{p} = a_0T^{-2} + \alpha T$, yielding $a_0=27 \mu$JK/mol. The $a_0$ value is then fixed with the $a_1$ as the only variable in fitting the low-temperature $C_\mathrm{p}$ data with a nonzero field. We found the data can be well fitted with $a_1$ = 123~$\mu$J K/(mol T$^2$), giving $C_{\rm n}=(27+123H^2)/T^2$ $\mu$J/(molK). This nuclear heat capacity was subtracted from all $C_\mathrm{p}$ data measured with the application of a magnetic field [Fig.~2(f) in the main text].

We note the coefficient $a_1$ is larger than expected from the nominal composition of the sample. In principle,  the Zeeman energy contribution to the heat capacity can be calculated as $a_1 = \sum n_I\hbar^2\gamma_{\rm I}^2I(I+1)/3k_{\rm B}$. The summation is over all atoms in the chemical formula with a nuclear spin $I \neq 0$. $n_I$ is the number of the nuclei per mole, $\hbar$ is the Planck constant, $\gamma_{\rm I}$ is the gyromagnetic ratio, and $k_{\rm B}$ is the Boltzmann constant. Based on the nominal chemical formula of the deuterated sample, one would expect $a_1 = $ 57~$\mu$J K/(mol T$^2$). Indeed, we found the fitted value of $a_1$ = 123~$\mu$J K/(mol T$^2$) would correspond to a H:D ratio close to 8:6, rather than all 14 H atoms are replaced by D. We suspect this is due to partially rehydrogenation  whereby H$_2$O replaces some of the D$_2$O during the crystal growth, leading to the discrepancy between the nominal and experimental determined $a_1$ values.

\begin{figure}[t]
	\centering
	\includegraphics[width=0.75\columnwidth]{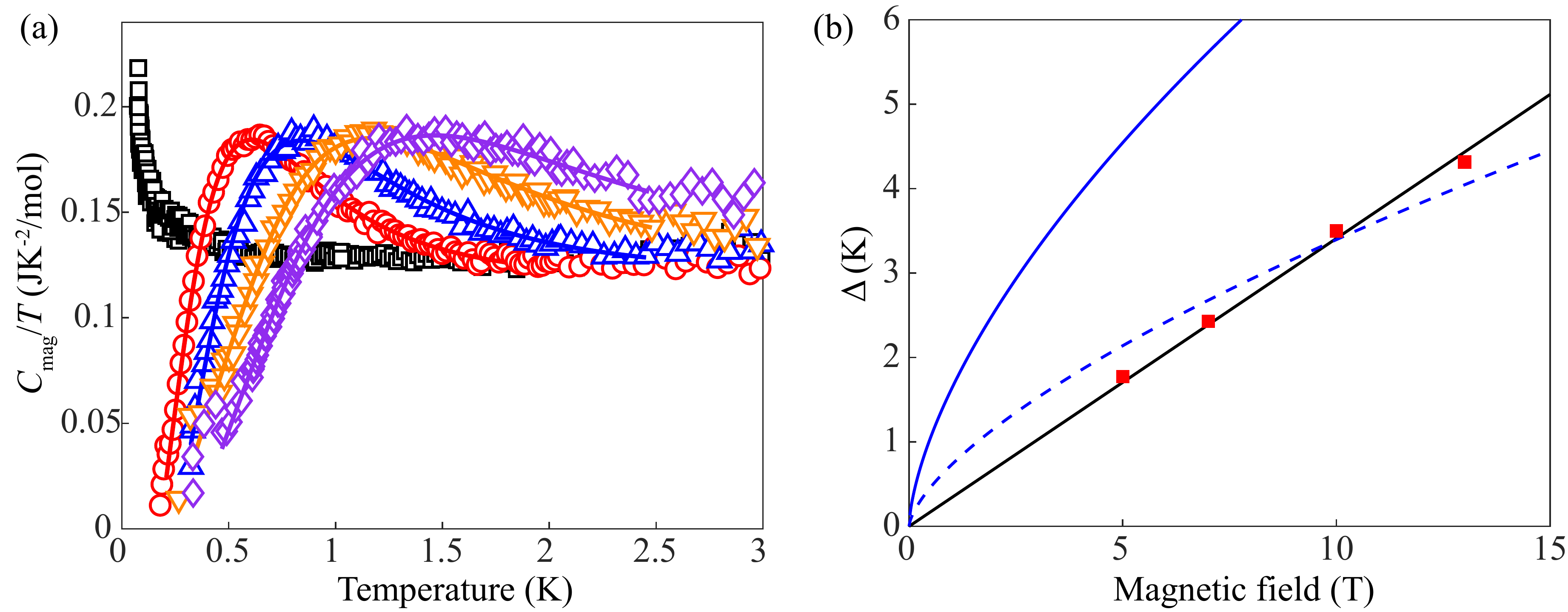}
	\caption{(a) Result of fitting the simple two-level model (lines) described below to experimental $C_{\rm mag}/T$ data (points). (b) Field dependence of the gap (squares) deduced from the fits in (a), as well as a best fit to a linear model (black line), a best fit to the SG model (dotted blue line), and the size of the gap predicted by the SG model using the experimentally-determined value of the two-fold staggered field (solid blue line).}
	\label{schott}
\end{figure}

\subsection{Fitting to obtain the gap size.} 

The temperature-dependence of the heat capacity of a gapped system will have a broad Schottky-like hump whose shape and position depends on the size of the gap and the degeneracy and distribution of excited states averaged across the Brillouin zone. In the manuscript the magnetic heat capacity is fitted with a model derived from the SG theory. While we stress that the SG model is not adequate to quantitatively account for many features of our data, there are still strong similarities between our system and that of the SG materials, particularly the staggered $g$-tensors and the richness of the excitation spectrum observed by ESR. Therefore the best possible expression that currently exists to reliably estimate the size of the gap from the heat capacity of our system is that derived from the SG model. 

Nevertheless, in order to show that the linear field dependence of the gap is not an artefact of using the SG heat-capacity expression, we also fit our data at fixed fields to a model consisting of a two-level system (Schottky anomaly) plus a constant $C/T$ term that accounts for the Tomonaga-Luttinger liquid behaviour observed at high temperatures. This is certainly not the ideal model to use to extract a reliable estimate of absolute size of the gap. This is because, while the heat capacity will be strongly influenced by the gap between the ground state and lowest energy excited state, any higher energy states will also have a significant effect. Such higher energy states are observed in the ESR spectra, but are not taken into account in this simple model. Thus we would expect the absolute size of the gap extracted in this way to be an overestimate. 

This is what we observe in Fig.~\ref{schott}: panel (a) shows the result of fitting this two-level model to our data and it is seen that the form of the data is reasonably well-described by the model. The points in panel (b) shows the size of the gap and its evolution in field. This procedure illustrates several issues: (i) the form of the heat capacity data are well described by a generic gapped model; (ii) the size of the gap extracted using this model, which is expected to be a overestimate of the actual gap, is still smaller across the measured field range than that predicted by the SG model using the experimentally-determined size of the two-fold staggered field in our material (solid blue line); and (iii) the gap thus extracted has a linear field-dependence (black line) and cannot be described by the SG model even with a staggered field much smaller than that measured (dotted blue line).

\section{Pulsed-field magnetization}

Pulsed-field magnetization experiments used a compensated-coil susceptometry technique, described in~\cite{PAG_NJP}. Fields were provided by the 65~T short-pulse and 100~T multi-shot magnets at NHMFL, Los Alamos. The susceptometer was placed within a $^4$He cryostat providing temperatures down to 1.6~K. Magnetic field was measured by integrating the voltage induced in a ten-turn coil calibrated by observing the de Haas-van Alphen oscillations of the belly orbits of the copper coils of the susceptometer. To create the traces shown in Fig.~\ref{Cp_and_M}(b), data below 65~T taken using the 65~T short pulse magnet were combined with data above 40~T taken using the 100~T multi-shot magnet. Both experiments made use of the same sample and susceptometer.

\section{Spin-wave expansion}

\begin{figure}[t]
	\centering
	\includegraphics[width=0.8\columnwidth]{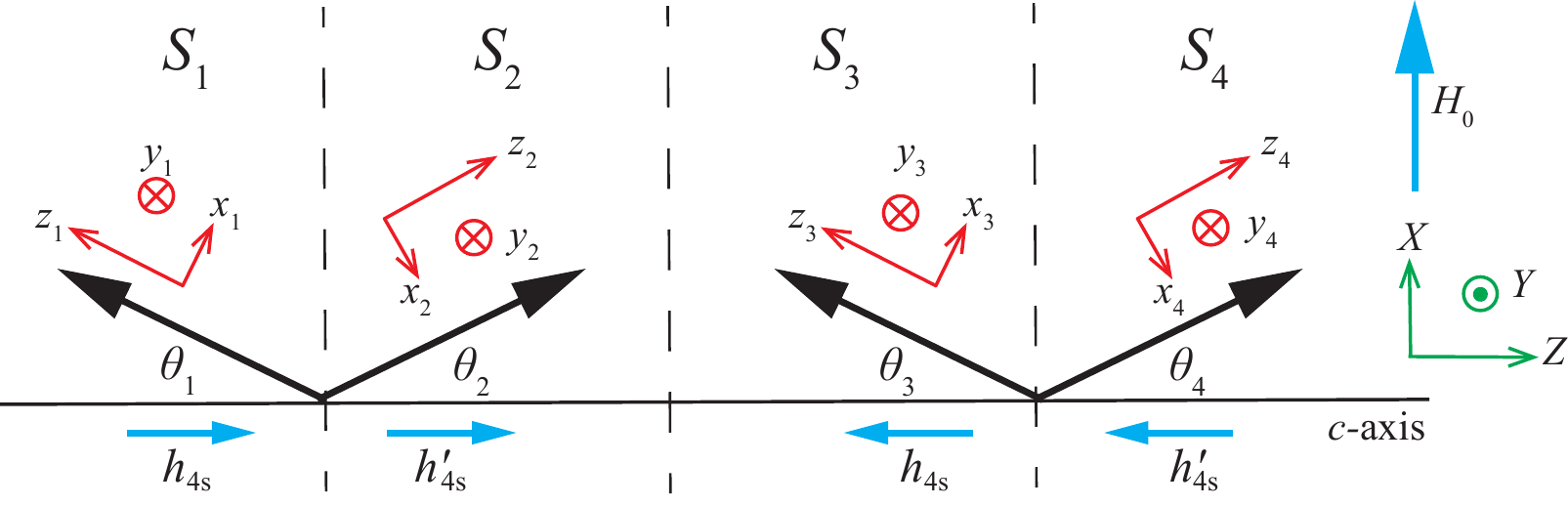}
	\caption{Schematic figure showing the $XZ$ canted ground state and the staggered fields for [Cu(pym)(H$_2$O)$_4$]SiF$_6$$\cdot$H$_2$O. The plot shows the crystallographic unit cell of Cu(pym), which includes four spins named as $S_1 \sim S_4$. The black arrows represents the spins for Cu(II) while the cyan arrows illustrating the arrangement of the staggered fields. The global and local coordination systems are labelled by green and red colours, respectively.}
	\label{fig1}
\end{figure}

We present a calculation for the excitation gap based on the standard spin-wave approximation for the $XZ$ canted structure with four-fold staggered fields $h_\mathrm{4s}$ and ${h_\mathrm{4s}^\prime}$. In the following calculation, we define the $XYZ$ frame as the laboratory frame. The $c$ axis of the chain is parallel to $Z$ and a external field $H_0$ is applied in the $X$ direction. The $x_iy_iz_i$ frames represent the local coordination systems for the $i$th \CuII ion where $z_i$ axes are defined by the direction of the $i$th spin (Fig.~\ref{fig1}). With an external field applied perpendicular to the chain, the four-fold staggered fields are parallel to the chain propagation direction ($c$-axis). This is the situation illustrated in Fig.~\ref{fig1} (also see Fig.\,3(b) in the main text). The magnitudes of staggered fields, $h_\mathrm{4s}$ and ${h_\mathrm{4s}^\prime}$, depend on the explicit form or the staggered $g_methrm{4s}$ and can be treated as twin independent parameters. However, as long as the field is applied within the \textit{XY} plane, the resultant four-fold staggered field should always be parallel to the \textit{Z} axis. The canted antiferromagnetic configuration of the spins, as well as the polarities of stagger fields, are illustrated in Fig.~\ref{fig1}. The canted angles, $\theta_n$ ($n = 1$ to 4), are determined by both the external field $H_0$ and the staggered fields. Therefore, Eq.\,3 in the main text does not hold rigorously due to the possibility that $\theta_1 \not= \theta_2 \not= \theta_3 \not= \theta_4$. The energy per unit cell of the canted antiferromagnet state is:
\begin{equation}
\label{classicalEnergy}
\begin{aligned}
E = &-JS^2[\cos{(\theta_1+\theta_2)}+\cos{(\theta_2+\theta_3)}+\cos{(\theta_3+\theta_4)}+\cos{(\theta_4+\theta_1)}]\\
&-H_0S[\sin{\theta_1}+\sin{\theta_2}+\sin{\theta_3}+\sin{\theta_4}]\\
&+h_\mathrm{4s} S(\cos{\theta_1}-\cos{\theta_3})-{h_\mathrm{4s}^\prime} S(\cos{\theta_2}-\cos{\theta_4})
\end{aligned}
\end{equation}
This energy is minimized for $\partial{E}/\partial{\theta_n} = 0$, which gives:
\begin{equation}
\label{minEnergy}
\begin{aligned}
\frac{\partial{E}}{\partial{\theta_1}} = JS^2[\sin{(\theta_1+\theta_2)}+\sin{(\theta_1+\theta_2)}]-H_0S\cos{\theta_1}-h_\mathrm{4s} S\sin{\theta_1} = 0\\
\frac{\partial{E}}{\partial{\theta_2}} = JS^2[\sin{(\theta_1+\theta_2)}+\sin{(\theta_2+\theta_3)}]-H_0S\cos{\theta_2}+{h_\mathrm{4s}^\prime} S\sin{\theta_2} = 0\\
\frac{\partial{E}}{\partial{\theta_3}} = JS^2[\sin{(\theta_2+\theta_3)}+\sin{(\theta_3+\theta_4)}]-H_0S\cos{\theta_3}+h_\mathrm{4s} S\sin{\theta_3} = 0\\
\frac{\partial{E}}{\partial{\theta_4}} = JS^2[\sin{(\theta_3+\theta_4)}+\sin{(\theta_4+\theta_1)}]-H_0S\cos{\theta_4}-{h_\mathrm{4s}^\prime} S\sin{\theta_4} = 0
\end{aligned}
\end{equation}

The staggered fields, $h_\mathrm{4s}$ and ${h_\mathrm{4s}^\prime}$, are much smaller than $H_0$. The canted angles, $\theta_n$, are mostly determined by $H_0$ while the staggered fields lead to a small correction to $\theta_n$. Therefore, the canted angles can be written as $\theta_n = \theta + \delta\theta_n$ where $\theta = \arcsin{(H_0/4JS)}$ is the canted angle induced by $H_0$. $\delta\theta_n$'s are the small corrections caused by the staggered fields and $\theta \gg \theta_n$. Hence, Eq.\,\ref{minEnergy} can be written as:
\begin{equation}
\label{minEnergy2}
\begin{aligned}
&16J^2S^2\delta\theta_1+(8J^2S^2-H^2)(\delta\theta_2+\delta\theta_4) = 2h_\mathrm{4s} H_0\\
&16J^2S^2\delta\theta_2+(8J^2S^2-H^2)(\delta\theta_1+\delta\theta_3) = -2{h_\mathrm{4s}^\prime} H_0\\
&16J^2S^2\delta\theta_3+(8J^2S^2-H^2)(\delta\theta_2+\delta\theta_4) = -2h_\mathrm{4s} H_0\\
&16J^2S^2\delta\theta_4+(8J^2S^2-H^2)(\delta\theta_1+\delta\theta_3) = 2{h_\mathrm{4s}^\prime} H_0
\end{aligned}
\end{equation}

Solving Eq.~\ref{minEnergy2} leads to a simple correlation between $\delta\theta_n$ and the staggered fields that:
\begin{equation}
\label{Angles}
\begin{aligned}
&\delta\theta_1 = \frac{h_\mathrm{4s} H_0}{8J^2S^2}\\
&\delta\theta_2 = -\frac{{h_\mathrm{4s}^\prime} H_0}{8J^2S^2}\\
&\delta\theta_3 = -\frac{h_\mathrm{4s} H_0}{8J^2S^2}\\
&\delta\theta_4 = \frac{{h_\mathrm{4s}^\prime} H_0}{8J^2S^2}\\
\end{aligned}
\end{equation}
At this point, it is more convenient to write the staggered field in terms of $\delta \theta_n$. Comparing Eq.~\ref{Angles} with Fig.~\ref{fig1}, we can get that the staggered field on the $nth$ site, $h_n$, is:
\begin{equation}
\label{fieldangle}
h_n = (-1)^{n+1}\frac{8J^2S^2}{H_0}\delta\theta_n.
\end{equation}
The effective spin Hamiltonian for Cu(pym) in the laboratory frame ($XYZ$ in Fig~\ref{fig1}) can be written as:
\begin{equation}
\label{Hamiltonian}
\hat{\cal{H}} = \sum_{n} J(\hat{S}^X_n\hat{S}^X_{n+1} + \hat{S}^Y_n\hat{S}^Y_{n+1} + \hat{S}^Z_n\hat{S}^Z_{n+1} )-H_0\hat{S}_n^{X}+(-1)^n\frac{8J^2S^2}{H_0}\delta \theta_nS_n^{Z}.
\end{equation}

The first term in Eq.\,\ref{Hamiltonian} represents the antiferromagnetic interactions. The second  and third terms are the Zeeman interactions due to the external field $H_0$ and the staggered fields $h_n$, respectively. $\hat{S}_n^X$, $\hat{S}_n^Y$ and $\hat{S}_n^Z$ correspond to the spin operators in the laboratory frame. They are related to the spin operators in the rotating frame ($\hat{S}_n^x$, $\hat{S}_n^y$ and $\hat{S}_n^z$) in the following way:
\begin{equation}
\label{rotation}
\begin{aligned}
&\hat{S}_n^X = (-1)^{(n+1)}\hat{S}_n^x\cos{\theta_n} + \hat{S}_n^z\sin{\theta_n}\\
&\hat{S}_n^Y = -\hat{S}_n^y\\
&\hat{S}_n^Z = (-1)^n\hat{S}_n^z\cos{\theta_n} + \hat{S}_n^x\sin{\theta_n}
\end{aligned}
\end{equation}
In the rotating frames, the ground state of the chain corresponds to the $n$th spin being parallel to $z_n$. By substituting Eq.\,\ref{rotation} into Eq.\,\ref{Hamiltonian} the Hamiltonian in the rotating frame can written as:
\begin{equation}
\label{Hamiltonian1}
\begin{aligned}
\hat{\cal{H}} = \sum_{n} &J[-\cos{(\theta_n+\theta_{n+1})}(\hat{S}_n^x\hat{S}_{n+1}^x+\hat{S}_n^z\hat{S}_{n+1}^z)+\hat{S}_n^y\hat{S}_{n+1}^y+(-1)^{n+1}\sin{(\theta_n+\theta_{n+1})}(\hat{S}_n^x\hat{S}_{n+1}^z-\hat{S}_n^z\hat{S}_{n+1}^x)]\\ 
&+ H_0[(-1)^n\cos{\theta_n}\hat{S}_n^x-\sin{\theta_n}\hat{S}_n^z]\\
&+ (-1)^n\frac{8J^2S^2}{H_0}\delta \theta_n[(-1)^n\cos{\theta_n}\hat{S}_n^z + \sin{\theta_n}\hat{S}_n^x],
\end{aligned}
\end{equation}

The spin operators can be written via bosonic operators using the Holstein-Primakoff transformation. The leading order expansion for a spin pointing in the $z$-direction is:
\begin{equation}
\label{HPtransform}
\begin{aligned}
&\hat{S}_n^x = \sqrt{\frac{S}{2}}(a_n^\dagger+a_n)\\
&\hat{S}_n^y = i\sqrt{\frac{S}{2}}(a_n^\dagger-a_n)\\
&\hat{S}_n^z = S-a_n^\dagger a_n
\end{aligned}
\end{equation}
Substituting Eq.\,\ref{HPtransform} into Eq.\,\ref{Hamiltonian1}, the Hamiltonian can be written as $\hat{\cal{H}} = \hat{\cal{H}}_0+\hat{\cal{H}}_1+\hat{\cal{H}}_2+\ldots + O(H_0) + O(h_\mathrm{4s},{h_\mathrm{4s}^\prime})$. $\hat{\cal{H}}_n = O(S^{2-n/2})$ is the field independent part whereas $O(H_0)$ and $O(h_\mathrm{4s},{h_\mathrm{4s}^\prime})$ correspond to the contributions due to the applied field $H_0$ and the four-fold staggered fields, respectively. By substituting Eq.\,\ref{HPtransform} into Eq.\,\ref{Hamiltonian1}, we get:

\begin{equation}
\label{Hexpansion}
\begin{aligned}
\hat{\cal{H}}_0 &= \sum_n -JS^2\cos{(\theta_n+\theta_{n+1})}\\
\hat{\cal{H}}_1 &= \sum_n (-1)^{(n+1)}J\frac{S^{3/2}}{2^{1/2}}[\sin{(\theta_{n-1}+\theta_n)}+\sin{(\theta_n+\theta_{n+1})}](a_n^\dagger+a_n)\\
\hat{\cal{H}}_2 &= \sum_n JS[\cos{(\theta_n+\theta_{n+1})}+\cos{(\theta_{n-1}+\theta_n)}]a_n^\dagger a_n\\
&\hspace{5em}-\frac{JS}{2}[\cos{(\theta_n+\theta_{n+1})}+1](a_n^\dagger a_{n+1}^\dagger+a_na_{n+1})+\frac{JS}{2}[1-\cos{(\theta_n+\theta_{n+1})}](a_n^\dagger a_{n+1}+a_n a_{n+1}^\dagger)\\
O(H_0) &= \sum_n -H_0\sin{\theta_n}S + H_0(-1)^n\cos{\theta_n}\sqrt{\frac{S}{2}}(a_n^\dagger + a_n) + H_0\sin{\theta_n}a_n^\dagger a_n\\
O(h_\mathrm{4s},{h_\mathrm{4s}^\prime}) &= \sum_n [\frac{8J^2S^2}{H_0}\delta \theta_n(S - a_n^\dagger a_n) + (-1)^n\sqrt{\frac{S}{2}}\frac{8J^2S^2}{H_0}\delta \theta_n\sin{\theta_n}(a_n^\dagger - a_n)]
\end{aligned}
\end{equation}

Eq.\,\ref{Hexpansion} includes the first three leading orders in the exchange interaction part ($\hat{\cal{H}}_0$, $\hat{\cal{H}}_1$ and $\hat{\cal{H}}_2$) and all orders in $H_0$, $h_\mathrm{4s}$  and ${h_\mathrm{4s}^\prime}$. $\hat{H}_0$ is a constant. Taking into account the constraints for canted angles (Eq.\,\ref{minEnergy}) and replacing $\theta_n$ with $\theta + \delta\theta_n$, the Hamiltonian can be written as (to the order of $S$ in spin-wave expansion):

\begin{equation}
\label{H3}
\begin{aligned}
\hat{\cal{H}} = \sum_n &2JS\cos{2\theta}a_n^\dagger a_n+H_0\sin{\theta}a_n^\dagger a_n\\
&-\frac{JS}{2}(1 + \cos{2\theta})(a_n^\dagger a_{n+1}^\dagger+a_na_{n+1})+\frac{JS}{2}(1-\cos{2\theta})(a_n^\dagger a_{n+1}+a_n a_{n+1}^\dagger)\\
&+\frac{JS}{2}\sin{2\theta}(\delta\theta_n+\delta\theta_{n+1})(a_n^\dagger a_{n+1}^\dagger+a_na_{n+1} + a_n^\dagger a_{n+1} + a_n a_{n+1}^\dagger) - \frac{8J^2S^2}{H_0}\cos{\theta}\delta \theta_na_n^\dagger a_n\\
&+\frac{JS}{4}\cos{2\theta}(\delta\theta_n + \delta\theta_{n+1})^2(a_n^\dagger a_{n+1}^\dagger + a_n a_{n+1} + a_n a_{n+1}^\dagger + a_n^\dagger a_{n+1})\\
& [\frac{16J^2S^2 - H_0^2}{H_0^2}\sin{\theta}\delta \theta_n^2 - \frac{JS}{2}\cos{2\theta}(2\delta\theta_n^2 + \delta\theta_{n-1}^2 + \delta\theta_{n+1}^2)]a_n^\dagger a_n\\
\end{aligned}
\end{equation}

The first four terms of Eq.\,\ref{H3} correspond to the spin wave expansion for a uniform antiferromagnetic chain in the presence of a external field $H_0$ without any staggered field. The fifth and sixth terms are the linear contribution of $h_\mathrm{4s}$ and/or ${h_\mathrm{4s}^\prime}$ while the remaining two parts are proportional to $h_\mathrm{4s}^2$ and/or ${h_\mathrm{4s}^\prime}^2$. By Fourier transforming Eq.\,\ref{H3}, it can be found that the fifth and sixth terms vanish because that the values of $\delta\theta_n$ oscillate between $\pm h_\mathrm{4s} H_0/8J^2S^2$ (or $\pm {h_\mathrm{4s}^\prime} H_0/8J^2S^2$). Therefore, the Hamiltonian is:
\begin{equation}
\label{HFinal}
\begin{aligned}
\hat{\cal{H}} = \sum_k &[2JS\cos{2\theta} + H_0\sin{\theta} - \frac{H_0^2(16J^2S^2 - 3H_0^2)(h^2_\alpha + h^2_\beta)}{1024J^5S^5}\\
&\hspace{5em} + (JS-JS\cos{2\theta} + \frac{H_0^2(8J^2S^2 - H_0^2)(h^2_\alpha + h^2_\beta)}{1024J^5S^5})\cos{k}]a_k^\dagger a_k\\
&- [\frac{JS}{2}(1+\cos{2\theta}) - \frac{H_0^2(8J^2S^2 - H_0^2)(h^2_\alpha + h^2_\beta)}{2048J^5S^5}]\cos{k}(a_k^\dagger a_{-k}^\dagger + a_k a_{-k}).
\end{aligned}
\end{equation}

For clarity, we define the following functions:
\begin{equation}
\label{coefficients}
\begin{aligned}
A(k) &= [2JS\cos{2\theta} + H_0\sin{\theta} - \frac{H_0^2(16J^2S^2 - 3H_0^2)(h^2_\alpha + h^2_\beta)}{1024J^5S^5} + (JS-JS\cos{2\theta} + \frac{H_0^2(8J^2S^2 - H_0^2)(h^2_\alpha + h^2_\beta)}{1024J^5S^5})\cos{k}]\\
B(k) &= [\frac{JS}{2}(1+\cos{2\theta}) - \frac{H_0^2(8J^2S^2 - H_0^2)(h^2_\alpha + h^2_\beta)}{2048J^5S^5}]\cos{k},
\end{aligned}
\end{equation}
and $\hat{\cal{H}} = \sum_k A(k)a_k^\dagger a_k - B(k)(a_k^\dagger a_{-k}^\dagger + a_k a_{-k})$. By performing the following Bogoliubov transformation:
\begin{equation}
\label{Bogoliubov}
\begin{aligned}
a_k = \cosh{\psi_k}\alpha_k + \sinh{\psi_k}\alpha_{-k}^\dagger\\
a_{-k}^\dagger = \sinh{\psi_k}\alpha_k + cosh{\psi_k}\alpha_{-k}^\dagger
\end{aligned}
\end{equation}
with $\tanh{2\psi_k} = 2B(k)/A(k)$, the Hamiltonian (Eq.\,\ref{HFinal}) can be diagonalised that:
\begin{equation}
\label{Hdiag}
\hat{\cal{H}} = \sqrt{A^2(k) - 4B^2(k)}\alpha^\dagger_k\alpha_k.
\end{equation}

which gives a single band in the paramagnetic Brillouin zone. We can equivalently fold the dispersion relation into the antiferromagnetic Brillouin zone, $-\pi/2 < k < \pi/2$. This gives us two branches with dispersion relations:

\begin{equation}
\label{energybands}
\begin{aligned}
E_\pm = \{&[2JS\cos{2\theta} + H_0\sin{\theta} - \frac{(h_\mathrm{4s}^2 + {h_\mathrm{4s}^\prime}^2)H_0^2\cos{2\theta}}{32J^3S^3}
\pm [JS(1-\cos{2\theta}) + \frac{(h_\mathrm{4s}^2 + {h_\mathrm{4s}^\prime}^2)H_0^2\cos{2\theta}}{64J^3S^3}]\cos{k}]^2\\ 
&- [JS(1+\cos{2\theta}) - \frac{(h_\mathrm{4s}^2 + {h_\mathrm{4s}^\prime}^2)H_0^2\cos{2\theta}}{64J^3S^3}]^2\cos^2{k}\}^{1/2}
\end{aligned}
\end{equation}

In the absence of staggered fields, $E_- = 0$ and Eq.\,\ref{energybands} gives the gapless excitation expected for uniform antiferromagnetic chains. Non-zero four-fold staggered fields lead to a excitation gap ($\Delta$) with the magnitude of the gap being: 
\begin{equation}
\label{gap}
\begin{aligned}
\Delta = \sqrt{\frac{(h_\mathrm{4s}^2 + {h_\mathrm{4s}^\prime}^2)H_0^2}{J^2S^2}\times\frac{(16J^2S^2 - H_0^2)(8J^2S^2 - H_0^2)}{1024J^4S^4}}
\end{aligned}
\end{equation} 

Eq.\,\ref{gap} shows that the excitation gap is proportional to $\sqrt{(h_\mathrm{4s}^2 + {h_\mathrm{4s}^\prime}^2)H_0^2/J^2S^2}$. The staggered fields are much smaller than the applied field $H_0$ and the antiferromagnetic interactions $JS$. Therefore, comparing with the excitation gap in Cu benzoate~\cite{OshikawaPRL1997a,AffleckPRB1999} ($\Delta \propto \sqrt{hJS}$) calculated with similar spin-wave expansion technique, the gap in Cu(pym) is expected to be much smaller. We note that by taking into account 1D critical fluctuation, the power-law behavior of the gap in Cu benzoate is changed from $h^{1/2}$ to $h^{2/3}$. We expected similar corrections should also be applied to our calculations, modifying the prediction of the field induced gap.

\end{document}